# INITIAL DEMONSTRATION OF 9-MHz FRAMING CAMERA RATES ON THE FAST DRIVE LASER PULSE TRAINS*

A. H. Lumpkin[#], D. Edstrom Jr., and J. Ruan
Fermi National Accelerator Laboratory, Batavia, IL 60510 USA

*Abstract*

We report the configuration of a Hamamatsu C5680 streak camera as a framing camera to record transverse spatial information of green-component laser micropulses at 3- and 9-MHz rates for the first time. The latter is near the time scale of the ~7.5-MHz revolution frequency of the Integrable Optics Test Accelerator (IOTA) ring and its expected synchroton radiation source temporal structure. The 2-D images are recorded with a Gig-E readout CCD camera. We also report a first proof of principle with an OTR source using the linac streak camera in a semi-framing mode.

## INTRODUCTION

Although beam centroid information at the MHz-micropulse-repetition rate has routinely been achieved at various facilities with rf BPMS, the challenge of recording beam size information at that rate is more daunting due to limitations in data-transfer rates. This is also near the time scale of the 7.5-MHz revolution frequency of the Integrable Optics Test Accelerator (IOTA) ring being constructed at the Fermilab Accelerator Science and Technology (FAST) Facility [1]. To simulate the expected IOTA optical synchrotron radiation (OSR) source temporal structure, we have used the green component of the FAST drive laser [2]. This is normally set at 3 MHz, but has also been run at up to 9 MHz. To circumvent the need to readout the 2D images in less than a few microseconds, we have configured our Hamamatsu C5680 streak camera as a framing mode camera using a slow vertical sweep plugin unit with the dual axis horizontal sweep unit. A two-dimensional array of images sampled at the MHz rate can then be displayed on the streak tube phosphor and recorded by the CCD readout camera at up to 10 Hz.

Demonstrations of the tracking of the beam size and position of consecutive green micropulses are shown, although there are gaps in the displayed pulse train for a given trigger delay. As an example, by using the 10 microsecond vertical sweep with the 100 microsecond horizontal sweep ranges, 49 of the 300 micropulses at 3 MHz are displayed for a given trigger delay. The whole pulse train dynamics are shown by recording only six sets of images with the appropriate stepped delays. Spatial resolutions of better than 15 microns seem possible for beam profiling and would be even better for beam centroids. Example 2D image arrays with profiling examples will be presented.

## EXPERIMENTAL ASPECTS

Two main aspects of the experiment are the drive laser as the source of a visible-light, 3-MHz pulse train and the Hamamatsu streak camera configured as a framing camera.

### The Drive Laser & FAST Electron Accelerator

The drive laser is based on a Calmar seed laser, consisting of a Yb-doped fiber laser oscillator running at 1.3 GHz that was then divided down to 81.25 MHz before amplification through a set of fiber amplifiers as shown in Fig.1. The seed output of 81.25 MHz packets of 1054 nm infrared (IR) laser is then reduced to the desired pulse train frequency (nominally 3 MHz) with a Pockels cell before selection of the desired pulse train with two additional Pockels cells and amplification through a series of YLF crystal-based single pass amplifiers (SPAs) and a Northrup-Grumman amplifier, which nominally yields 50 µJ of IR per pulse before the two frequency-doubling crystal stages generate the green and then the UV components with a total nominal efficiency of 10% [2]. The pulse train selected is between a single pulse per machine cycle (nominally 1 Hz) and 1 ms (3000 pulses at the nominal 3 MHz pulse train frequency) [2]. The UV drive laser pulse train is used to generate an electron pulse in the FAST IOTA electron injector, an SRF-based linear accelerator tested thus far to 50 MeV [3].

In the initial framing camera studies we observed the green component remaining from UV-conversion at 3 and 9 MHz pulse train frequencies with the laser lab streak camera, but we have also recently applied the principle to optical transition radiation (OTR) from an Al-coated Si substrate foil with subsequent transport to a beamline streak camera.

### The Streak Camera Systems

Commissioning of the streak camera system was facilitated through a suite of controls centered around ACNET, the Fermilab accelerator controls network. This suite includes operational drivers to control and monitor the streak camera as well as Synoptic displays to facilitate interface with the driver. Images from the readout cameras, Prosilica 1.3 Mpixel cameras with 2/3" format, may be analyzed both online with a Java-based ImageTool and an offline MATLAB-based ImageTool processing program [4,5]. Bunch-length measurements using these techniques have been reported previously from the A0 Facility [6] and FAST first commissioning at 20 MeV [7].

___________________
*Work supported under Contract No. DE-AC02-07CH11359 with the United States Department of Energy.
#lumpkin@fnal.gov

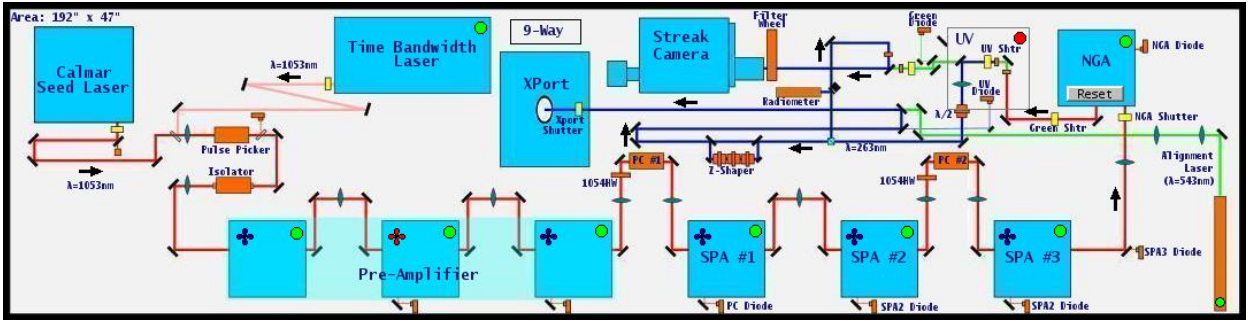

Figure 1: Schematic of the FAST drive laser optical layout showing the seed laser, pre-amplifier, SPAs, and the streak camera.

The streak camera stations each consisted of a Hamamatsu C5680 mainframe with S20 PC streak tube and can accommodate vertical sweep plugin units and either a horizontal sweep unit or a blanking unit. The UV-visible input optics allow the assessment of the 263-nm component as well as the amplified green component or IR components converted to green by a doubling crystal. We started the framing mode studies by replacing the M5675 synchroscan unit with its resonant circuit tuned to 81.25 MHz with the M5677 slow vertical sweep unit (5-ns to 1-ms ranges) [8]. The M5679 dual axis plugin which provides a horizontal sweep with selectable ranges had already been installed for the previous dual sweep synchroscan tests as indicated in Fig. 2. A second set of deflection plates in the streak tube provides the orthogonal deflection for the slower time axis in the 100-ns to 10-ms regime. These plates are driven by the dual-axis sweep unit which was also commissioned during previous studies. The C5680 Gate Trigger In was connected to a DG535 TTL output with the correct gating time.

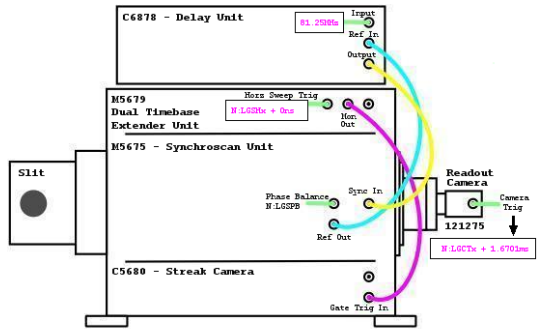

Figure 2: The Laser Lab streak camera wiring diagram. N:LGSHx and N:LGCTx are ACNET names for beam-synchronized VME-based timers.

## EXPERIMENTAL RESULTS

### Initial Green Component studies: 3 MHz

The basic principles are illustrated with the Laser Lab Streak Camera in Figs. 3-5. Fig. 3 shows the focus mode image that is constrained vertically by the entrance slits to the streak camera optics barrel, but in principle images the horizontal profile directly. In this case the vertically apertured image is about 8.1 pixels or 32 μm (sigma) while the horizontal size is 53 pixels or 212 μm.

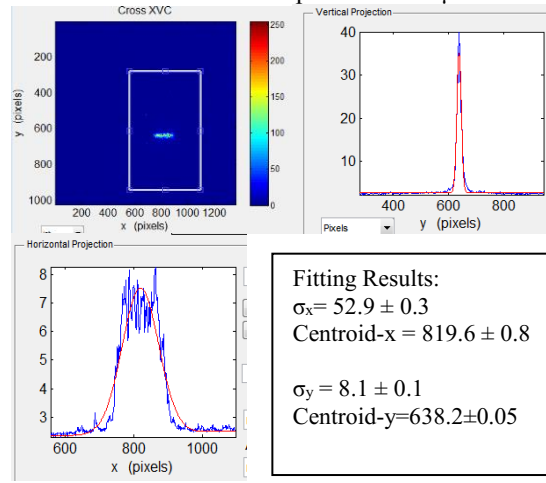

Figure 3: Focus mode image of streak camera with horizontal and vertical projections shown for the drive laser green component. Gaussian fitting results are shown.

In Fig. 4 we have applied a slow vertical sweep with a range of 10 μs to display the 20-micropulse-long pulse train. The vertical projected profiles on the ROI would show all 20 micropulses at a 3-MHz rate. Such an image could track bunch-by-bunch centroid motion as well, particularly in the horizontal plane perpendicular to the sweep direction.

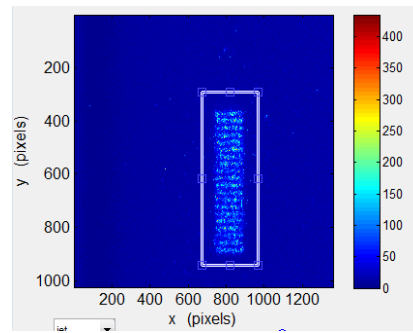

Figure 4: Vertical slow sweep image of 20 micropulses with 10-μs sweep range. Green at 3 MHz.

To obtain more pulse separation vertically, we reduce the coverage to 1 µs with a faster deflection as shown in Fig. 5. We also add the horizontal deflection with a span of 100 µs. In this case, we lengthened the macropulse to 310 micropulses, and we display 46 of them at a time for a set delay trigger. We can cover all 310 pulses with selected, stepped trigger delays in 6 images. Each micropulse image can be processed for profile and position.

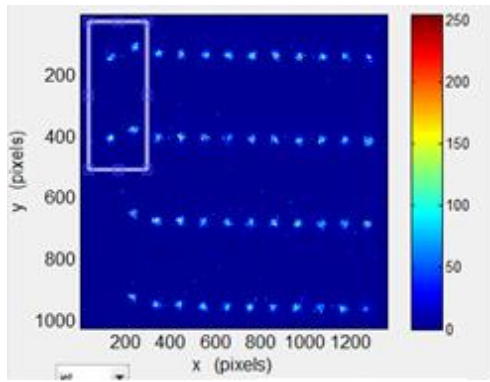

Figure 5: Framing camera mode (1 µs x 100 µs). 46/310 micropulses are shown. 3 MHz green component.

### Green Component: 9 MHz

To simulate the IOTA ring OSR source's 7.5-MHz revolution frequency, the drive laser table Pockels cells were adjusted to switch out a 9-MHz IR pulse train which then went to the first doubling crystal. In Fig. 6 we show 9 consecutive micropulses in vertical columns of 1 µs range and 13 columns that sampled the pulse train of the green component. The fits of the x and y projections in the indicated ROI can then be processed.

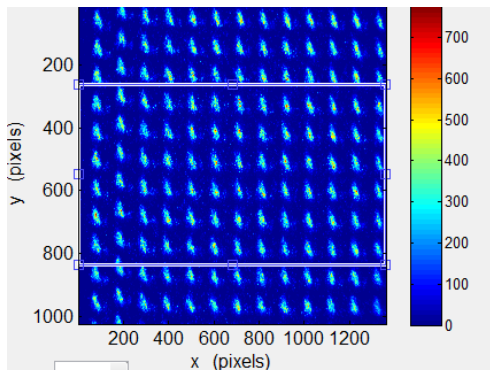

Figure 6: Framing camera mode (1 µs x 100 µs) with Green component at 9 MHz.

### Linac streak camera with OTR

The Electron Beamline Streak Camera is installed in an optical enclosure outside of the beamline enclosure with transport of OTR from the instrumentation cross at beamline location 121 (X121) as described elsewhere in this conference [9]. The all-mirror transport allows us to minimize the chromatic temporal dispersion effects for bunch length measurements. The same transport can be used for the framing-mode tests. In this case, we only used the horizontal sweep to separate the micropulses at 3 MHz. The input image was apertured by the vertical slit and horizontally to provide 135 µm by 25 µm sigma-x,y sizes, respectively, for the demonstration shown in Fig. 7. The spatial resolution for the system is in the 10- to 15- µm range with an effective calibration factor of 6.6 µm/pixel. The initial micropulse charge of 300 pC thus was reduced by the aperture in the camera image. This proof of principle with OTR can be applied to detecting higher order mode (HOM) dipole mode effects in the e-beam pulse train or turn-by-turn OSR effects in IOTA.

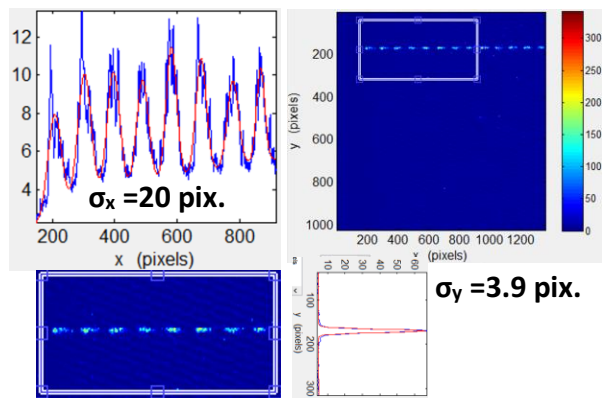

Figure 7: Semi-framing-camera mode with X121 OTR from the electron beam pulse train at 3 MHz.

### SUMMARY

In summary, we have described a series of results using the laser lab Hamamatsu streak camera in framing mode to track beam size and position at 3 and 9 MHz. We also have our first result from the X121 OTR station showing that we can apply the technique on individual e-beam micropulses such as in support of HOM effect studies. We calculate that the OSR sources in IOTA [10] will be brighter turn by turn than the OTR in the electron beamline so we have established the proof of principle for that application as well.

### ACKNOWLEDGMENTS

The authors acknowledge the support of A. Valishev, D. Broemmelsiek, N. Eddy, and R. Dixon, all at Fermilab.
This research is dedicated *in memoriam* to Helen Edwards.